\documentclass[longbibliography,groupedaddress,showpacs,showkeys,amssymb,eqsecnum,aps,nofootinbib,superscriptaddress]{revtex4-2}
\usepackage[]{graphicx}
\usepackage[]{graphics}
\usepackage{amsmath}
\usepackage{epsf}                                                                                           
\usepackage{color}                   
\usepackage{verbatim}                                                                         
\usepackage{hyperref}

\newcommand{\be}{\begin{equation}}

\newcommand{\ee}{\end{equation}}

\begin{document}                                                                                            
\author{F. T. Brandt}
\email{fbrandt@usp.br}
\affiliation{Instituto de F\'{\i}sica, Universidade de S\~ao Paulo, S\~ao Paulo, SP 05508-090, Brazil}
\author{J. Frenkel}
\email{jfrenkel@if.usp.br}
\affiliation{Instituto de F\'{\i}sica, Universidade de S\~ao Paulo, S\~ao Paulo, SP 05508-090, Brazil}
\author{D. G. C. McKeon}
\email{dgmckeo2@uwo.ca}
\affiliation{
Department of Applied Mathematics, The University of Western Ontario, London, Ontario N6A 5B7, Canada}
\affiliation{Department of Mathematics and Computer Science, Algoma University, Sault Ste.~Marie, Ontario P6A 2G4, Canada}
\author{G.  S.  S.  Sakoda}   
\email{gustavo.sakoda@usp.br}
\affiliation{Instituto de F\'{\i}sica, Universidade de S\~ao Paulo, S\~ao Paulo, SP 05508-090, Brazil}

\title{Loop corrections in a solvable UV-finite model and its effective field theory}

\date{\today}

\begin{abstract}
We examine some features of the non-renormalizability induced through the use of low-energy effective Lagrangians in loop diagrams, in the context of a scalar model which is ultraviolet finite and partially soluble. In this framework, one can directly demonstrate the mechanism leading
to the non-renormalizability of the effective theory. This behavior is generated by approximations that are applicable at low energies but are generally inappropriate for evaluating loop diagrams that contain virtual high-energy particles. However, it is explicitly shown that one can match the results obtained in the renormalized effective theory with those found in the full theory at low energy. We argue that the infrared sectors of these theories are inherently similar, independently of the matching procedure. A closed-form expression is obtained, to leading order in the energy expansion, for the complete effective Lagrangian at all orders in perturbation theory. The model may be useful to clarify certain aspects of realistic, but more complex, effective field theories.
\end{abstract}

\pacs{11.10Kk,04.60Kz}
\keywords{effective field theories, soluble models}

\maketitle

\section{Introduction}\label{sec1} 
 
Effective field theories are approximations which include the proper degrees of freedom that describe phenomena occurring at accessible energies but disregard the substructure of the underlying theories at higher energies \cite{weinberg:book2005,Penco:2020kvy,Cohen:2019wxr,duncan:book2012}.
This procedure works because under certain conditions, the heavy particles decouple from the low-energy physics (modulo a renormalization of the parameters of the theory) \cite{applequist}. 
Effective theories are most useful when the coupling constants with negative mass dimensions involve a mass scale that is much larger than the typical energy of the process being studied. One of the best-known examples is the effective Fermi theory, which is obtained in the low-energy limit of the electroweak theory \cite{Fermi:1934hr}. Another interesting instance is the effective Euler-Heisenberg Lagrangian, which arises in the low energy limit of the photon-photon scattering amplitude in QED \cite{Euler:1936oxn,Dunne:2012vv}. A further example is the effective theory of the pion-nucleon interaction, which describes certain properties of QCD at low energies \cite{Leutwyler:1999mz}. Other important aspects and applications of effective field theories for the Standard Model are discussed in references \cite{donoghue_golowich_holstein_2014,Dobado:1997jx,Burgess2021}.

     Any effective theory may include, apart from renormalizable interactions, an infinite number of non-renormalizable interactions. In particular, there are no renormalizable interactions in Einstein's general relativity \cite{Einstein:1916vd}, which is expected to be a low energy effective theory of some fundamental quantum theory. There is an extensive literature on this subject, as can be seen in the reviews \cite{Burgess:2003jk,Carlip:2015asa,Donoghue:2017,Donoghue:2017pgk} 
and references cited therein. Accordingly, the gravitational interactions that occur at low energies may be described by Einstein-Hilbert Lagrangian. Using the tree vertices deduced from this effective Lagrangian, one can perform perturbative calculations of loop diagrams in quantum gravity \cite{tHooft:1974bx,Goroff:1986th,vandeVen:1991gw}. But such contributions require an infinite number of counter-terms to cancel all the ultraviolet divergences in this effective field theory \cite{Gomis:1995jp,PhysRevD.100.026018}.

In this work, we examine some features of such a non-renormalizability, which can be induced by the use of effective Lagrangians beyond the tree approximation. To this end, we study in section \ref{sec2} a solvable one-dimensional scalar field theory involving 
two interacting scalar fields, which is ultraviolet finite. In this context, we clarify the procedure leading to the non-renormalizability of the effective theory. This feature occurs due to approximations that hold at low energy, but are generally unsuitable for the evaluation of loop diagrams containing internal high-energy particles. Nevertheless, one can include an infinite number of counter-terms to cancel all non-renormalizable divergences in the effective theory. We compute the scattering amplitude at tree level and the self-energy at one loop for light particles in the renormalized effective model. One can match the results found in this theory with the corresponding ones obtained, respectively, at one loop and at two loops in the full theory at low energy. In this way, the effective field theory becomes consistent and can be used to predict other low-energy effects. In Sec. \ref{sec3} we analyze the effective Lagrangian, up to next to leading order in the energy expansion. To leading order, we deduce an exact result for the complete effective Lagrangian to all orders in perturbation theory (Eq. \eqref{e19}). In section \ref{sec4}, a basic argument is given for the agreement between the infrared sectors of the effective and the full theory, 
that occurs irrespective of the matching procedure.
We conjecture that somewhat similar features may possibly also arise in the usual formulation of quantum gravity, which relies upon the effective Einstein-Hilbert Lagrangian.
The scattering amplitude at higher orders is briefly examined in the Appendix \ref{appB}.
In the Appendix \ref{appAa} we present a derivation of the static effective action.

\section{A solvable scalar model}\label{sec2}

We will consider here a workable one-dimensional scalar field theory that involves
two interacting scalar fields, described by the Lagrangian
\be\label{e1}
{\cal L}(\phi,\psi) = \frac 1 2 \left(\partial_t\phi\partial_t\phi-\lambda^2 \phi^2\right)+\partial_t\psi^\star \partial_t\psi-m^2 \psi^\star\psi -ig\phi\left(\psi^\star\partial_t\psi-\psi\partial_t\psi^\star\right)
\ee
where the real scalar field $\phi$ is much lighter than the complex field $\psi$ ($\lambda\ll m$). The mass dimension of the coupling constant $g$ is $3/2$, so that the model is super-renormalizable.

    The only diagrams that appear to be, by power counting, logarithmically ultraviolet divergent are tadpole graphs that have a single external light field. But in consequence of charge conjugation invariance, the Green functions with an odd number of external light fields vanish, so that this theory is actually ultraviolet finite.

    The model leads to a Feynman diagrammatic representation of perturbative processes that is somewhat similar to that in scalar QED. However, unlike the case in QED where the photon-photon scattering amplitude cannot be evaluated in closed form \cite{r23}, 
in this simple theory one can obtain analytical results for Feynman amplitudes with external light fields.

\subsection{The scattering amplitude}

In this model, one can exactly evaluate the scattering amplitudes of the light field coming from the diagrams shown in Fig. (1). There are three other diagrams that differ only by the arrow direction in the closed-loop, that yield the same contribution. These graphs may also occur as part of a larger diagram, where some light field lines become virtual and do not satisfy the condition $k\ll m$, as will be discussed next. It is, therefore, useful to evaluate the diagrams with  
\be\label{e2}
k^2_i\neq\lambda^2; \;\;\;\; k_1+k_2+k_3+k_4=0.
\ee
\begin{figure}[b]
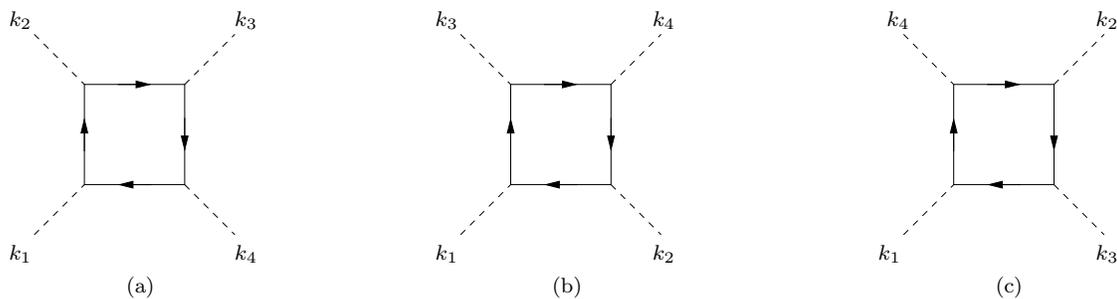

\input fig1_frenkel2020.pspdftex
\caption{One-loop diagrams for the scattering amplitude of the light field. The full line denotes the virtual heavy particle.}\label{fig1}
\end{figure}
For reasons of symmetry, it is convenient to treat all four light particles as being incoming. Using the theorem of residues to evaluate the integral over the internal energy, together with the fact that only three external energies are independent, we then obtain for the four-point amplitude  the analytical expression 
\be\label{e3}
M(k_1,k_2,k_3)=-\frac{4g^4}{m}\left[
\frac{1}{(k_1+k_2)^2-4 m^2+i\epsilon}+
\frac{1}{(k_2+k_3)^2-4 m^2+i\epsilon}+
\frac{1}{(k_3+k_1)^2-4 m^2+i\epsilon}
\right] .
\ee
We note that this explicit result exhibits poles at the threshold energy $2m$ for pair production. At low energies such that $|k_1|$, $|k_2|$, $|k_3|\ll m$, in the approximation corresponding to an energy 
expansion up to the next to leading order, the above equation reduces  to the expression
\be\label{e4}
M_{0}(k_1,k_2,k_3)\approx\frac{g^4}{m^3}\left[3+
\frac{(k_1+k_2)^2+(k_2+k_3)^2+(k_3+k_1)^2}{4 m^2}
\right] .
\ee
We remark that Eq. \eqref{e3} is well behaved when the energies become very large. If this result is used for boxes that are sub-diagrams in higher-loop graphs, like those shown in Fig. 2, one can see that this leads to contributions that are rapidly converging for large values of $k_3$.
On the other hand, Eq. \eqref{e4} involves terms that are quadratic in the energies. If this effective 4-point vertex is used in effective loop graphs like those shown in Fig. 3,  these terms will induce spurious linear ultraviolet divergences, coming from the region $k_3\rightarrow\infty$ (see the next subsection).
The reason for this bad ultraviolet behavior is due to the inappropriate use of Eq. \eqref{e4} in the region where $|k_3| \gg m$. In this region, the low-energy expression \eqref{e4} is invalid and one must use the correct result \eqref{e3} in order to ensure a good high-energy behavior of the amplitudes.
\begin{figure}[t]
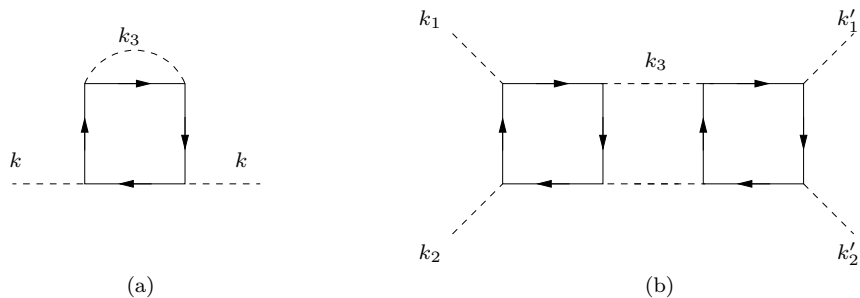

\input fig2_frenkel2020.pspdftex 
\caption{Examples of higher-order Feynman diagrams with light external particles.}\label{fig2}
\end{figure}
\begin{figure}[b]
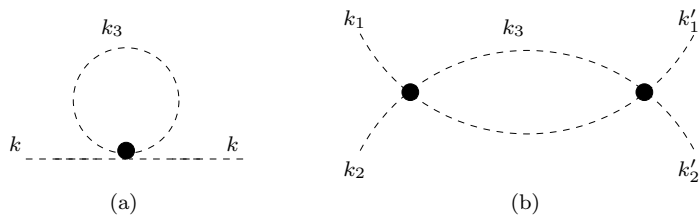

\input fig3_frenkel2020.pspdftex 
\caption{Examples of effective one-loop graphs. The small blob denotes the effective low-energy 4-point vertex of the light field.}\label{fig3}
\end{figure}

\subsection{The self-energy function}
 The one-loop diagram for the self-energy of the light field is shown in Fig. 4a. Performing the integration over the internal energy by using the residue theorem, we get the result
\be\label{e5}
\Pi^{(2)} = -\frac{g^2}{m}
\ee
which is independent of the external energy and real. An imaginary part does not occur at the threshold energy for pair production, $k  = 2m$, since at this point the cubic vertices vanish.

\begin{figure}[b]
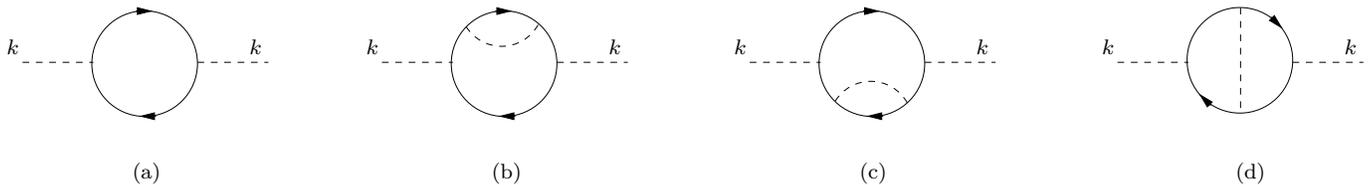

\input fig4_frenkel2020.pspdftex 
\caption{Loop diagrams for the self-energy of the light field.}\label{fig4}
\end{figure}

The evaluation of the one-particle irreducible self-energy in the full theory at two loops, shown in Figs. (4b), (4c) and (4d), is more involved as one has to take into account the bubble and the vertex sub-diagrams.
Carrying out the integrations over the energies of the virtual particles in 
these figures, one obtains for the self-energy at two-loops the analytical result
\be\label{e6}
\Pi^{(4)}(k)=-\frac{g^4}{m^2}\left[
\left(\frac{2 m}{\lambda}+1\right)\frac{1}{k^2-(2 m+\lambda)^2+i\epsilon}-\frac{1}{4 m\lambda}\right]
\ee
which is ultraviolet finite. For the reason given following Eq. \eqref{e5}, this expression has a pole only at the threshold energy for three particle production. Assuming that $\lambda\ll m$ and $k\ll m$,
one can expand the exact result \eqref{e6} up to terms of order $k^2$, getting
\be\label{e7}
\Pi^{(4)}(k)\approx\frac{g^4}{4 m^3}\left[\frac 1 \lambda
\left(3+\frac{k^2}{2 m^2}\right)-\frac 1 m -
\frac{3\lambda}{2 m^2}+\frac{k^2-\lambda^2}{4 m^3}\right]. 
\ee
Let us now compare  the low-energy  result  \eqref{e7} with the one obtained by using the effective 4-point vertex \eqref{e4}. One then finds that the graph in Fig (3a), where the effective vertex is given by the expression \eqref{e4} with $k_1= - k_2 = k$, yields the contribution
\be\label{e8}
\Pi^{(4)}_{ef}(k) = \frac{g^4}{2 m^3}\left[\frac{1}{2 \lambda}
\left(3+\frac{k^2 +\lambda^2}{2 m^2}\right)-\frac{\Lambda}{4 m^2}\right], 
\ee
where we have introduced in the last term a Pauli-Villars regulator $\Lambda$. In contrast to \eqref{e7}, this contribution exhibits a spurious linear divergence, which was induced by the improper use of the effective low-energy 4-point vertex \eqref{e4}. As we have pointed out, such an approximation is invalid at high values of the energy $k_3$.
One may cancel the linear ultraviolet divergence in Eq. \eqref{e8} by introducing a suitable counter-term $C$ which can be generated by a renormalization of the mass of the light particle in the Lagrangian \eqref{e1}
\be\label{e10n}
\lambda^2 = \lambda_r^2 + \frac{g^2}{m} C
= \lambda_r^2 - \frac{g^4}{4 m^4}
\left(\frac{\Lambda}{2 m}-1 -\frac{2\lambda}{m}\right).
\ee
In this way, using Eq. \eqref{eq31aa}
together with Eq. \eqref{e8}, we obtain for $k$, $\lambda \ll m$ the following result
\be\label{e10}
\Pi^{(4)\, a}_{ ef}(k) = \frac{g^4}{4 m^3}\left[\frac{1}{\lambda}
  \left(3+\frac{k^2}{2 m^2}\right)-\frac 1 m - \frac{3\lambda}{2 m^2}
  \right].
\ee
Comparing Eq. \eqref{e10} with the relation \eqref{e7} obtained at low energy in the full theory, we can see that one still must match the last term in this relation. To this end, we must also consider a contribution from the wave-function renormalization of the light field in the Lagrangian \eqref{e1}  given by $\phi = Z^{1/2} \phi_r$, with
\be\label{e13n}
Z = 1 + \frac{g^4}{16 m^6}.
\ee
Using Eq. \eqref{eq31aa} to order $g^4$,
this gives the contribution 
\be\label{e13}
\Pi^{(4)\, b}_{ ef}(k) = \frac{1}{16}\frac{g^4}{m^6}
\left(k^2-\lambda^2\right).
\ee
Together with Eq. \eqref{e10}, this yields a result which agrees with that given in Eq. \eqref{e7}. There is no coupling-constant renormalization at order $g^4$. Thus, one can match the  result obtained for the self-energy in the renormalized effective theory at one-loop, to the one calculated at two-loops in the full theory at low energy.

We note here that the infrared singular contributions, which occur in the limit $\lambda\rightarrow 0$ of the terms proportional to $1/\lambda$, are the same in both theories, independently of any matching 
\be\label{e11}
\Pi^{(4)}_{\lambda}(k) = \Pi^{(4)}_{ef,\,\lambda}(k) = 
\frac{g^4}{4 m^3}\left[\frac{1}{\lambda}
\left(3+\frac{k^2}{2 m^2}\right)\right]. 
\ee
This happens because  the infrared singular terms arise from the region where all internal energies are also small, in which case the low-energy approximations are adequate.

\section{The effective Lagrangian}\label{sec3}
Let us consider a local effective Lagrangian involving the light scalar field $\phi$, up to terms which are of next to leading order in the energy expansion. Using Eq. \eqref{e1}, the most general expression for such a Lagrangian may be written, up to order $g^4$, in the form 
\be\label{eq31}
{\cal L}_{ef}(\phi,\lambda) =
\frac 1 2\left(\partial_t\phi\partial_t\phi-\lambda^2\phi^2\right)
+l_0\frac{g^2}{2m} \phi^2
+ \frac{g^4}{8 m^3}\left(l_1\phi^4+\frac{l_2}{m^2}\phi^2\partial^2_t\phi^2
+\frac{l_3}{m^2}\phi^3\partial^2_t\phi\right),
\ee
where $l_0$, $l_1$, $l_2$ and $l_3$ are some dimensionless parameters
which have a perturbative expansion in terms of the dimensionless quantity $g^2/m^3$.

After a renormalization of the field $\phi$ and its mass $\lambda$, given by
\be\label{eq31a}
\phi = Z^{1/2} \phi_r;\;\;\;\; \lambda^2 = \lambda^2_r+\frac{g^2}{m} C,
\ee
the effective Lagrangian \eqref{eq31} may be written, up to order $g^4$, as 
\be\label{eq31aa}
{\cal L}_{ef}(\phi_r,\lambda_r) =
\frac Z 2\left[\partial_t\phi_r\partial_t\phi_r-\left(\lambda_r^2+\frac{g^2}{m} C\right)\phi_r^2\right]
+l_0\frac{g^2}{2m} Z\phi_r^2
+ \frac{g^4}{8 m^3}\left(l_1\phi_r^4+\frac{l_2}{m^2}\phi_r^2\partial^2_t\phi_r^2
+\frac{l_3}{m^2}\phi_r^3\partial^2_t\phi_r\right).
\ee
The parameters in Eq. \eqref{eq31aa} 
may be determined so as to bring 
agreement between the renormalized effective theory and the full theory at low energies. Thus, comparing the one-loop self-energy in the effective theory with the self-energy up to two loops in the full theory, fixes the values of $Z$ and $C$ as shown in Eqs. \eqref{e10n} and \eqref{e13n}, and determines the parameter $l_0$ to be $l_0=-1$.
Moreover, comparison of the tree-level scattering amplitude in the effective theory with the one-loop scattering amplitude in the exact theory at low energy, fixes to lowest order the parameters $l_1$, $l_2$ and $l_3$ to be $l_0=1$, $l_2=1/4$ and $l_3=0$.

We may obtain the complete effective Lagrangian
by integrating out the heavy field  $\psi$  in the full theory described by the Lagrangian \eqref{e1}
\be\label{e18}
\exp i \int dt \left[\frac 1 2\left(\partial_t\phi\partial_t\phi-\lambda^2\phi^2\right)
+{\cal L}_{ef}^{loop}(\phi)\right] =
\int {\cal D}\psi {\cal D}\psi^\star\exp i \int dt {\cal L}(\phi,\psi),
\ee
where ${\cal L}_{ef}^{loop}(\phi)$ is determined by the requirement that 
it vanishes when $g = 0$. ${\cal L}^{loop}_{ef}$ corresponds to the sum of all loop contributions with an 
arbitrary number of external light fields. This can be evaluated by using 
the  method given in \cite{Brown:1975bc}, which assumes that the fields are slowly varying. In the energy space, this condition corresponds to small values of $k$ such that $k \ll  m$. The integrations can be performed iteratively order by order in perturbation theory, leading to a rather involved result. The lowest order terms in the perturbative expansion yield the expression given in Eq. \eqref{eq31}, with the correct values of the parameters $l_i$. On the other hand, when $\phi=\varphi$ is a static field, one obtains a simple closed-form expression for the exact static effective Lagrangian to all orders in $g$ 
\be\label{e19}
{\cal L}_{ef}^{loop}(\varphi) = m - (m^2+g^2\varphi^2)^{1/2}.
\ee
We note that Eq. \eqref{e19} corresponds to the case where all external legs carry zero energy, so that  $-{\cal L}^{loop}_{ef}(\varphi)$ gives the effective potential. This is positive and bounded from below, which ensures the vacuum stability of the theory. A more direct derivation of Eq. \eqref{e19} may be obtained by noticing that in Eq. \eqref{e1}, one can make the following transformation of the  heavy field 
\be\label{e20}
\tilde\psi(t) = \psi(t)\exp i g \int_{-\infty}^t dt^\prime \phi(t^\prime).
\ee
In this way, we can write the Lagrangian \eqref{e1} in the alternative form 
\be\label{e21}
{\cal L}(\phi,\tilde\psi)=\frac 1 2\left(\partial_t\phi\partial_t\phi-\lambda^2\phi^2\right)
+\partial_t\tilde\psi^\star\partial_t\tilde\psi - m^2 \tilde\psi^\star\tilde\psi 
- g^2 \phi^2 \tilde\psi^\star\tilde\psi .
\ee
Using this Lagrangian, we integrate out the heavy field  $\tilde\psi$ in Eq. \eqref{e18}, to get
\be\label{e22}
\exp i \int dt {\cal L}^{loop}_{ef}(\phi) = {\det}^{-1}\left(
1+\frac{1}{\partial_t^2+m^2} g^2 \phi^2
\right)
=\exp - {\rm Tr}\left[\ln\left(1+\frac{1}{\partial_t^2+m^2} g^2 \phi^2\right)\right] ,
\ee
where we used the condition that ${\cal L}^{loop}_{ef}$ vanishes at $g=0$.
%


We assume that the field $\phi(t)$ varies alowly in time and reduces to the field $\varphi$ in the static limit. Doing first the time differentiations and taking next the limit $\phi\rightarrow\varphi$ in Eq. \eqref{e22}, one 
gets in energy space the relation 
\be\label{e23}
\exp i \int dt {\cal L}^{loop}_{ef}(\varphi)=\exp - \int dt \int_{-\infty}^{\infty}\frac{d q_0}{2\pi}
\ln\left(1+\frac{g^2\varphi^2}{m^2-q_0^2}\right).
\ee
As shown in Appendix \ref{appAa}, this result is applicable 
for static fields.

Expanding  the logarithm in a power series of $g^2$, one can integrate this series term by term 
\cite{gradshteyn}, yielding
\be\label{e24}
{\cal L}^{loop}_{ef}(\varphi)=m\sum_{k=1}^{\infty}\frac{(-1)^k}{2^k}\frac{(2k-3)!!}{k!}
\left(\frac{g^2\varphi^2}{m^2}\right)^k = m - \left(m^2+g^2\varphi^2\right)^{1/2},
\ee
which agrees with the expression given in Eq.  \eqref{e19}.

\section{Discussion}\label{sec4}
We have studied a solvable UV-finite scalar model and its associated effective field theory. In this simple framework, we aimed to clarify the mechanism leading to the nonrenormalizability of loop amplitudes, which arises by using the effective Lagrangian beyond the tree approximation.
This feature is induced by approximations that are applicable at low energies, but are generally incorrect for the evaluation of loop graphs involving virtual high-energy particles. 
One may include an infinite number of counter-terms to cancel all non-renormalizable ultraviolet
divergence in the effective theory, but this leads to a lack of predictivity at high energy. Nevertheless, we have shown explicitly that we can match the results obtained in the renormalized effective theory with the corresponding results found  at low energy in the full theory. Such a matching fixes all free parameters in the effective theory
which enables the use of this theory to evaluate other low-energy processes. This procedure
 works because the ultraviolet effects are local and can be absorbed into the local terms of the 
effective Lagrangian.  An explicit expression was obtained, to leading order in the energy expansion,
for the complete effective Lagrangian to all orders in perturbation theory (Eq. \eqref{e19}). 
This result yields the negative of the exact effective potential of the model, which is positive and has a lower bound, that assesses the vacuum stability of this theory.
    
There are important differences between the interaction
vertices in the effective theory and those in the full theory. A relevant point is that the low-energy approximations require that the energies of  all  particles emerging from the effective vertices, including the ones which may become internal in loop graphs, should be small. As we 
saw, this generates the non-renormalizability pointed out above. We have further analyzed this issue in the case of the self-energy to order $g^4$  and shown that the infrared singular terms in both theories are similar, independently of the matching scheme. This property was also explicitly verified in the scattering amplitude at higher orders (see Appendix \ref{appB}). A basic explanation of this behavior relies upon the fact that the infrared singularities arise from the emission and propagation of soft massless particles. In this
regime the internal loop energies are also 
small, in which case the low-energy approximations become reliable. Thus, the infrared sectors in the effective theory and in the full theory at low energy naturally turn out to be similar, regardless of the matching operation which is necessary in other sectors.
    
Furthermore, the non-analytic parts of the Feynman loop amplitudes, like  logarithmic terms,
are generally the same in both theories due to the fact that the S-matrix has just those singularities which are required by unitarity \cite{weinberg:book2005,Halter:1993kj,Eden:1966dnq}. In our  simple model there are no such terms, so that this property is manifestly satisfied.

   One may expect analogous features to appear as well in other effective field theories. Thus, it
   would be interesting to examine whether these properties could lead to a useful model for what might happen in a quantum theory of gravity based on the Einstein-Hilbert Lagrangian. Although the underlying theory has not yet been fully elucidated, the mechanism leading to the non-renormalizability of this effective theory may be rather similar. Here, the infrared divergences occur for soft gravitons coupled to other hard particles and their cancellation can be proved by the same method used in 
QED \cite{Weinberg:1965nx,Donoghue:1996mt}. Based on this fact along with the above reasoning, one may conjecture that the infrared sector of quantum general relativity could be somewhat analogous to that occurring at low energies in a fundamental renormalizable theory of gravity.
%
                                                         
\begin{acknowledgments}
We are much indebted to Prof.  J. C.Taylor for many valuable correspondences. We would also like to thank CNPq (Brazil) for financial support. 
\end{acknowledgments}  

\appendix

\section{Scattering amplitudes at higher orders}\label{appB}
Let us compare the scattering amplitudes in the full and the effective theories, shown respectively in Figs (\ref{fig2}b) and (\ref{fig3}b). Such amplitudes may be written in the form ($k_1+k_2=k^\prime_1+k^\prime_2$)
\be\label{a1}
\int \frac{dk_3}{2\pi i}
\frac{1}{k_3^2-\lambda^2+i\epsilon}
\frac{1}{(k_1+k_2+k_3)^2-\lambda^2+i\epsilon}
M(k_1,k_2,k_3) M(-k_1^\prime,-k_2^\prime,-k_3) .
\ee
In the full theory, $M(k_1,k_2,k_3)$ is given by Eq. \eqref{e3}. For simplicity, we consider the scattering 
in the special case where $k_1=k_2=k_1^\prime=k_2^\prime=k$. The result found by performing the $k_3$ integration is rather involved, but it simplifies in the low energy limit where
$k,\,\lambda\ll m$. In this case, we obtain
\be\label{a2}
-\frac{g^8}{m^6}\frac{1}{\lambda}\left[\left(\frac{9}{4}+\frac{3\lambda^2}{2m^2}\right)\frac{1}{k^2-\lambda^2+i\epsilon}+\frac{9}{4m^2}+\frac{2k^2+3\lambda^2}{4m^4}\right]+\frac{g^8}{m^9}\left(\frac{7}{8}+\frac{13k^2+9\lambda^2}{16 m^2}\right).
\ee
On the other hand, in the effective theory one employs the low-energy form given in Eq. \eqref{e4}. Then, after performing the $k_3$ integration, we obtain the result
\be\label{a3}
-\frac{g^8}{8m^9}\frac{\Lambda}{m}-\frac{g^8}{m^6}\frac{1}{\lambda}\left[\left(\frac{9}{4}+\frac{3\lambda^2}{2m^2}\right)\frac{1}{k^2-\lambda^2+i\epsilon}+\frac{9}{4m^2}
\right],
\ee
which has a spurious linear ultraviolet divergence.

Moreover, one must also take into account the corresponding contribution coming from the effective Lagrangian \eqref{eq31}, which is given by
\be\label{a4}
\frac{g^4}{m^3}\left[
3\Delta l_1+(8\Delta l_2-\Delta l_3)\frac{k^2}{2 m^2}\right] .
\ee
Requiring the sum of the expressions given by Eqs. \eqref{a3} and \eqref{a4} to match the result given in Eq. \eqref{a2}, yields the following corrections to the effective parameters
\be\label{a5}
\Delta l_1 = \frac{g^4}{m^6}\left(
\frac{1}{24}\frac{\Lambda}{m}+\frac{7}{24}-\frac{\lambda}{4 m}
+\frac{3\lambda^2}{16 m^2}\right) ,
\ee

\be\label{a6}
8 \Delta l_2 - \Delta l_3 = \frac{g^4}{m^6}\left(\frac{13}{8}-\frac{m}{\lambda}\right)
.
\ee

Although we have derived the above relations for a special configuration of the external energies, such relations hold in general since the effective parameters are independent of the energies. But in the general case, one would be able to fix $\Delta l_2$ and $\Delta l_3$ separately.

We note that the terms proportional to $1/(k^2-\lambda^2+i\epsilon)$ in Eqs. \eqref{a2} and \eqref{a3} are equal and do not require any matching. These terms yield at $k=0$, the leading infrared singularities which are proportional to $1/\lambda^3$. Such terms come from the region where all internal loop energies are also small (of order $\lambda$), in which case the low energy approximations are appropriate. This explains the equality of the above contributions. 

\section{The static effective Lagrangian}\label{appAa}
In the static case, one can see from Eqs. \eqref{e18} and \eqref{e21} that the Lagrangian ${\cal L}^{loop}_{ef}$ satisfies the relation
\be\label{nb1}
\exp i\int dt {\cal L}^{loop}_{ef}(\varphi) = N \int {\cal D}\tilde\psi
  {\cal D}\tilde\psi^{\star}\exp-i\int dt\tilde\psi^\star
  \left(\partial^2_t+m^2+g^2\varphi^2\right)\tilde\psi,
\ee
where $N$ is a normalization constant determined by the requirement that ${L}^{loop}_{ef}=0$ when $g=0$. Differentiating both sides of \eqref{nb1} with respect to $M^2\equiv m^2+g^2\varphi^2$, we get
\be\label{nb2}
\frac{\partial {\cal L}^{loop}_{ef}}{\partial M^2}= -N\langle\tilde\psi^\star(t)\tilde\psi(t)\rangle = iN G(t,t),
\ee
where $G(t,t)$ is the exact Green's function $G(t,t^\prime)$ evaluated at the coincidence point $t=t^\prime$. This Green's function satisfies the differential equation
\be\label{nb3}
\left(\partial^2_t+m^2+g^2\varphi^2\right)G(t,t^\prime)=\delta(t,t^\prime).
\ee
Using \eqref{nb3}, one can see that the Fourier-transformed Green's functions is given by
\be\label{nb4}
\tilde G(q_0) = \int dt e^{i q_0(t-t^{\prime})}G(t,t^\prime)=
\frac{1}{m^2+g^2 \varphi^2 - q_0^2}.
\ee
Thus, we find
\be\label{nb5}
G(t,t) = \int\frac{d q_0}{2\pi} \tilde G(q_0) = \int\frac{d q_0}{2\pi} 
\frac{1}{m^2+g^2 \varphi^2 - q_0^2}.
\ee
Substituting this result in Eq. \eqref{nb2} and integrating over $M^2$, we obtain the result
\be\label{nb6}
         {\cal L}^{loop}_{ef}(\varphi) = i N
         \int\frac{d q_0}{2\pi} \ln(m^2+g^2 \varphi^2 - q_0^2)=
i \int\frac{d q_0}{2\pi} \ln\left(1+\frac{g^2 \varphi^2}{m^2-q_0^2}\right),
\ee
which is equivalent to that given in Eq. \eqref{e23}.

\end{document}